\documentclass[preprint,12pt]{elsarticle}

\usepackage{graphicx}
\usepackage{amssymb}
\usepackage{booktabs}
\usepackage{color}
\usepackage{setspace}

\journal{Materials Research Bulletin}

\begin{document}

\begin{frontmatter}

\author[a,b,c]{L T Corredor\corref{cor1}}
\ead{ltcorredorb@df.ufpe.br}
\author[a,b]{J Albino Aguiar}
\author[c]{D A Land\'{i}nez T\'{e}llez}
\author[d]{P Pureur}
\author[d]{F Mesquita}
\author[c]{J Roa-Rojas}
\address[a]{Departamento de F\'{i}sica, Universidade Federal de Pernambuco, 50670-901, Recife-PE, Brasil}
\address[b]{Programa de P\'{o}s-Gradua\c{c}\~{a}o em Ci\^{e}ncias de Materiais-CCEN, Universidade Federal de Pernambuco, 50670-901, Recife-PE, Brasil}
\address[c]{Grupo de F\'{i}sica de Nuevos Materiales, Departamento de F\'{i}sica, Universidad Nacional de Colombia, Bogot\'{a} D.C., Colombia}
\address[d]{Instituto de F\'{i}sica, Universidade Federal do Rio Grande do Sul, 91501-970 Porto Alegre-RS, Brasil}

\title{Magnetic, electrical and structural properties of the new Re-doped ruthenocuprate Ru$_{1-x}$Re$_x$Sr$_2$GdCu$_2$O$_y$}

\begin{abstract}
Despite the discovery of new superconductors classes, high-Tc oxides continue to be a current topic, because of their complex phase diagrams and doping-dependant effects (allowing one to investigate the interaction between orbitals), as well as structural properties such as lattice distortion and charge ordering, among many others. Ruthenocuprates are magnetic superconductors in which the magnetic transition temperature is much higher than the critical superconducting temperature, making them unique compounds. With the aim of investigating the dilution of the magnetic Ru sub-lattice, we proposed the synthesis of the new Ru$_{1-x}$Re$_x$Sr$_2$GdCu$_2$O$_y$ ruthenocuprate-type family, adapting the known two-step process (double perovskite + CuO) by directly doping the double perovskite, thus obtaining the new perovskite compound Sr$_2$GdRu$_{1-x}$Re$_x$O$_y$, which represents a new synthesis process to the best of our knowledge. Our samples were structurally characterised through X-ray diffraction, and the patterns were analysed via Rietveld refinement. A complete magnetic characterisation as a function of temperature and applied field, as well as transport measurements were carried out. We discuss our results in the light of the two-lattice model for ruthenocuprates, and a relation between RuO$_{2}$ (magnetic) and CuO$_2$ (superconductor) sub-lattices can clearly be observed.
\end{abstract}

\begin{keyword}
A.superconductors \sep A.magnetic materials \sep D.electrical properties

\end{keyword}

\end{frontmatter}

\section{Introduction}

The discovery of superconductivity in the ruthenocuprates RuSr$_2$GdCu$_2$O$_8$ (Ru1212) and RuSr$_2$(R$_{1+x}$Ce$_{1-x}$)Cu$_2$O$_{10}$ (R=Sm, Eu and Gd) by Bauernfeind et al. in 1995 \cite{Bauernfeind}, and the report, two years later, by Felner et al. \cite{Felner} of the coexistence of superconductivity and magnetism in these compounds renewed the interest of both theoreticians and experimentalists, in the study of the interplay between superconductivity and magnetism \cite{Pimentel,Jurelo,Ruiz-Bustos,Ubaldini,McCrone,Nachtrab}.

The distinguishing characteristic of these compounds in comparison with other magnetic superconductors such as Chevrel phases \cite{Chevrel} or rare earth ternary borides \cite{Matthias} is the fact that the magnetic transition temperature is much higher than the critical superconducting temperature, making them unique materials. However, these compounds are extremely sensitive to the synthesis process, making their study difficult. In order to reduce the synthesis-dependent characteristics, a method for ruthenocuprate production involving the synthesis of double perovskites Sr$_2$LnRuO$_6$ (Ln=lanthanide) as precursor oxides, called the two-step process, was developed \cite{Zhigadlo,Bauernfeind2,Klamut}, opening at the same time a new research line dealing with these interesting lanthanide perovskites \cite{Doi,Triana,Triana2,Corredor}.

In accordance with these ideas, we proposed the synthesis of the new Re-doped ruthenocuprate Ru$_{1-x}$Re$_x$Sr$_2$GdCu$_2$O$_y$ in the interest of investigating the dilution of the magnetic Ru sub-lattice. After facing serious problems in obtaining this compound through the standard two-step process (adding the doping agent together with Sr$_2$GdRuO$_6$), we adapted it by directly doping  the double perovskite with Rhenium to get Sr$_2$GdRu$_{1-x}$Re$_x$O$_y$. This represents a completely new ruthenocuprate preparation technique to the best of our knowledge.

Felner et al. \cite{Felner} found via magnetic susceptibility and M\"{o}ssbauer spectroscopy that superconductivity seems to be confined to the CuO$_2$ planes whereas the magnetism is due to the Ru sublattice. In the present paper, we discuss our results in light of the two-lattice model for ruthenocuprates, where an alternating sequence of weakly ferromagnetic (RuO$_2$), insulating (SrO), and superconducting (CuO$_2$) sheets along the c-axis is formed (SIFIS), exhibiting an intrinsic Josephson effect.

\section{Experimental Details}

When RuSr$_2$GdCu$_2$O$_8$ was sintered for the first time, the sample did not have a single phase character, and the magnetic transition was observed at 138 K, which was attributed to the possible presence of SrRuO$_3$, known by its ferromagnetism. Impurities are caused by the volatility of RuO$_x$, which plays a fundamental role in the final Ru/Cu ratio, affecting the physical propeties of the samples. The appearance of perovskite impurities was minimized through a two-step method that consisted in producing the double perosvkite Sr$_2$GdRuO$_6$ (SGRO) and using it as a precursor powder together with CuO. This is currently the most accepted method for producing ruthenocuprate systems, and we applied it to our system, as explained above.

In our case, Sr$_2$GdRu$_{1-x}$Re$_x$O$_y$ (SGReO) with x= 0.00, 0.03, 0.06, 0.09, and 0.12 was prepared using the solid state reaction method, as described elsewhere \cite{Corredor}. SGReO perovskite powders were then mixed with CuO Aldrich powder (99,995 \%) (previously dried at 200 $^\circ$C for 24 hours), ground in an agate mortar, and then pressed into pellets of approximately 5 mm diameter and 1 mm thickness. The samples were thermally treated at 1050 $^\circ$C for 45 hours, with intermediate grindings.

Once obtained, an oxygen treatment was carried out in a quartz tube at 1 atm pressure, with the aim of optimizing the superconductor properties of the sample. Several lengths of times were tried, but for times longer than 120 hours in oxygen flux, the properties remained the same. The crystal structure was studied through X-ray powder diffraction, using a PanAlytical Pro diffractometer with Cu-k$\alpha$ radiation (1.5406 \AA) and PiXcel detection. The morphology and the qualitative chemical composition were studied with a FEI Quanta 200 ESEM, and its corresponding EDX accessory. The diffraction patterns were analyzed through Rietveld refinement using GSAS software. Magnetic measurements were carried out on a Quantum Design MPMS (Magnetic Properties Measurement System) SQUID magnetometer, and transport measurements were performed with a Quantum Design PPMS (Physical Properties Measurement System).

\section{Results and discussion}

\subsection{Structural characterization}

\begin{figure}[ht]
\centering
\includegraphics[width=12cm]{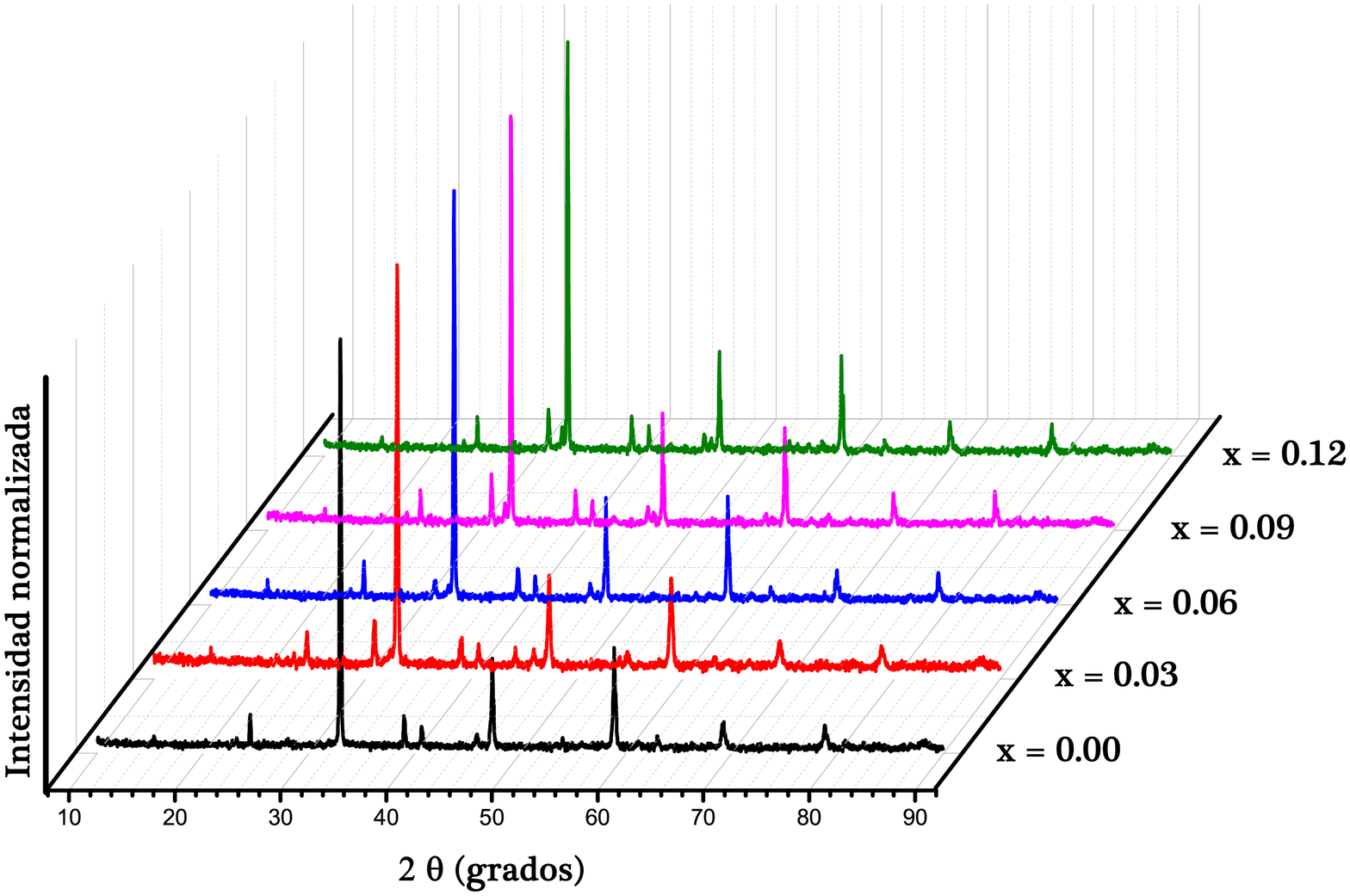}\quad
\includegraphics[width=12cm]{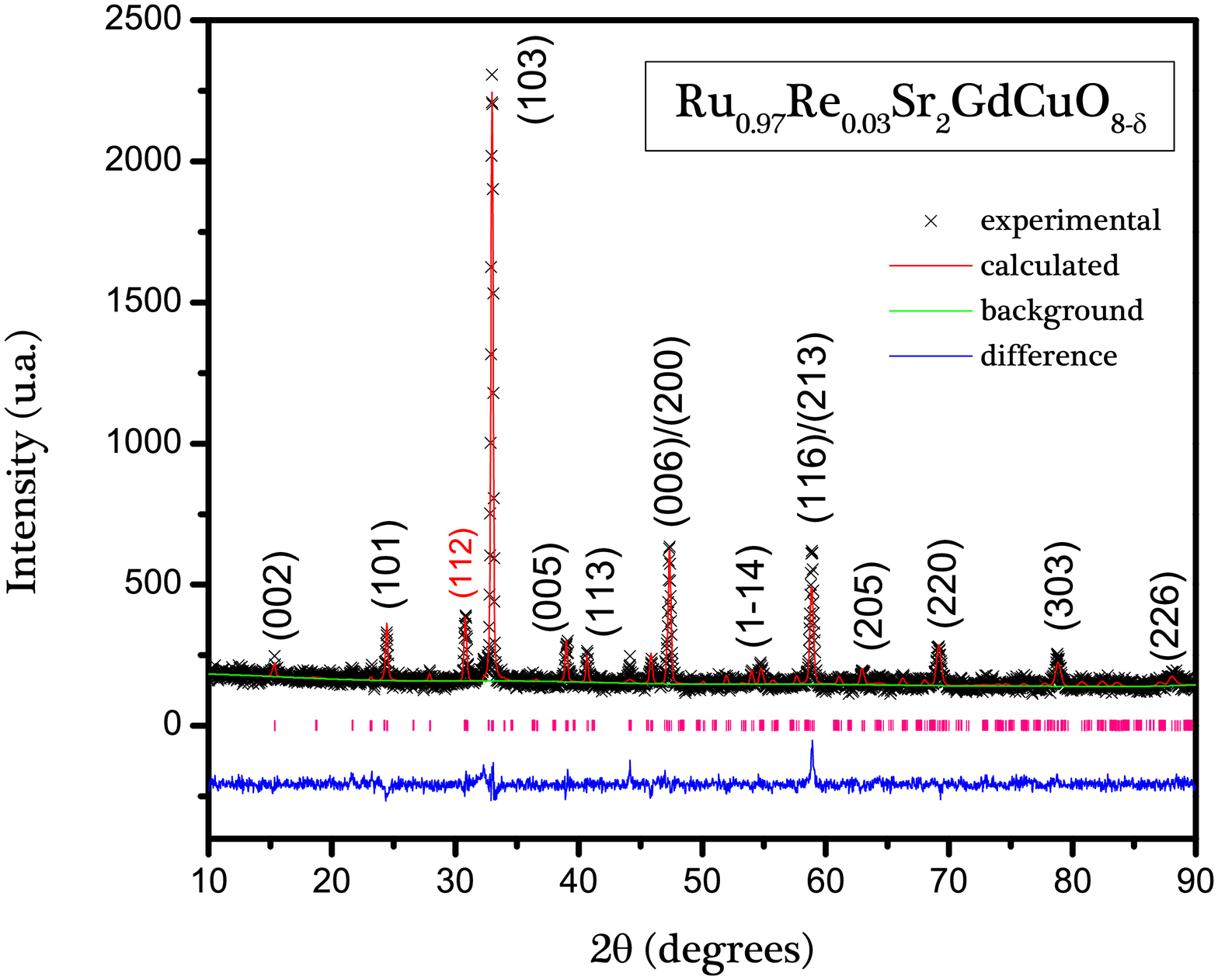}
\caption{}
\label{Fig.1}
\end{figure}

X-ray diffraction results are shown in Figure \ref{Fig.1}. All the samples are tetragonal, with space group P4mmm (\#123). Previous  reports on similar samples have found SrRuO$_3$ magnetic impurity, which makes the interpretation of the characterization results difficult \cite{Abatal}. However, with the two-step process previously described, that impurity does not appear. This has been attributed to the presence of pentavalent Sr$_2$GdRuO$_6$, which inhibits the formation of SrRuO$_3$, with tetravalent Ru, under oxidant conditions \cite{Braun}.

Rietveld refinement of X-ray diffraction patterns revealed that the samples are nearly single phase (see Figure \ref{Fig.1}(b) for results with x=0.03 sample). The presence of Sr$_2$GdRuO$_6$ was detected in the samples, in percentages between 0.3 and 2.2 \%: however, no Re phases were detected. This was confirmed through SEM imaging, where a compact structure with grains sizes between 2 and 10 $\mu$m were analyzed through EDX. This showed that no Re oxides were segregated, indicating, together with the diffraction results, that Re successfully enters into the structure.

Lattice parameters $a$ and $b$ show a slight tendency to increase, while $c$ diminishes to doping level x=0.09. At this point, the structure seems to re-accommodate, diminishing $a$ and $b$ and increasing $c$, as can be seen in Table \ref{Table1}.

The difference in the ionic ratio between Ru and Re is $\Delta r= 0.055$ if both of them are considered pentavalent, Ru$^{5+}$/Re$^{5+}$, and $\Delta r= 0.050$ if considered tetravalent, Ru$^{4+}$/Re$^{4+}$. To have an idea of the possible valences of Ru/Re in the compounds, nominal distances of Ru/Re-O bonds were calculated with the ionic ratios listed by Shannon \cite{Shannon}: for pentavalent Ru and Re ions this distance is approximately 1.965 \AA\, and 1.980 \AA\, respectively, and for tetravalent ions, 2.02 \AA\, and 2.03 \AA\, respectively. For our samples, the distances calculated from Rietveld refinement data have values between 1.972 and 2.032 \AA. This result suggests a mixed valence of Ru/Re ions in the compound, in the same way as reported for Ru-1212Gd type samples from NMR studies \cite{Kumagai,Liu,Yoshida}.

\subsection{Magnetic properties}

\begin{figure}[ht]
\centering
\includegraphics[width=11cm]{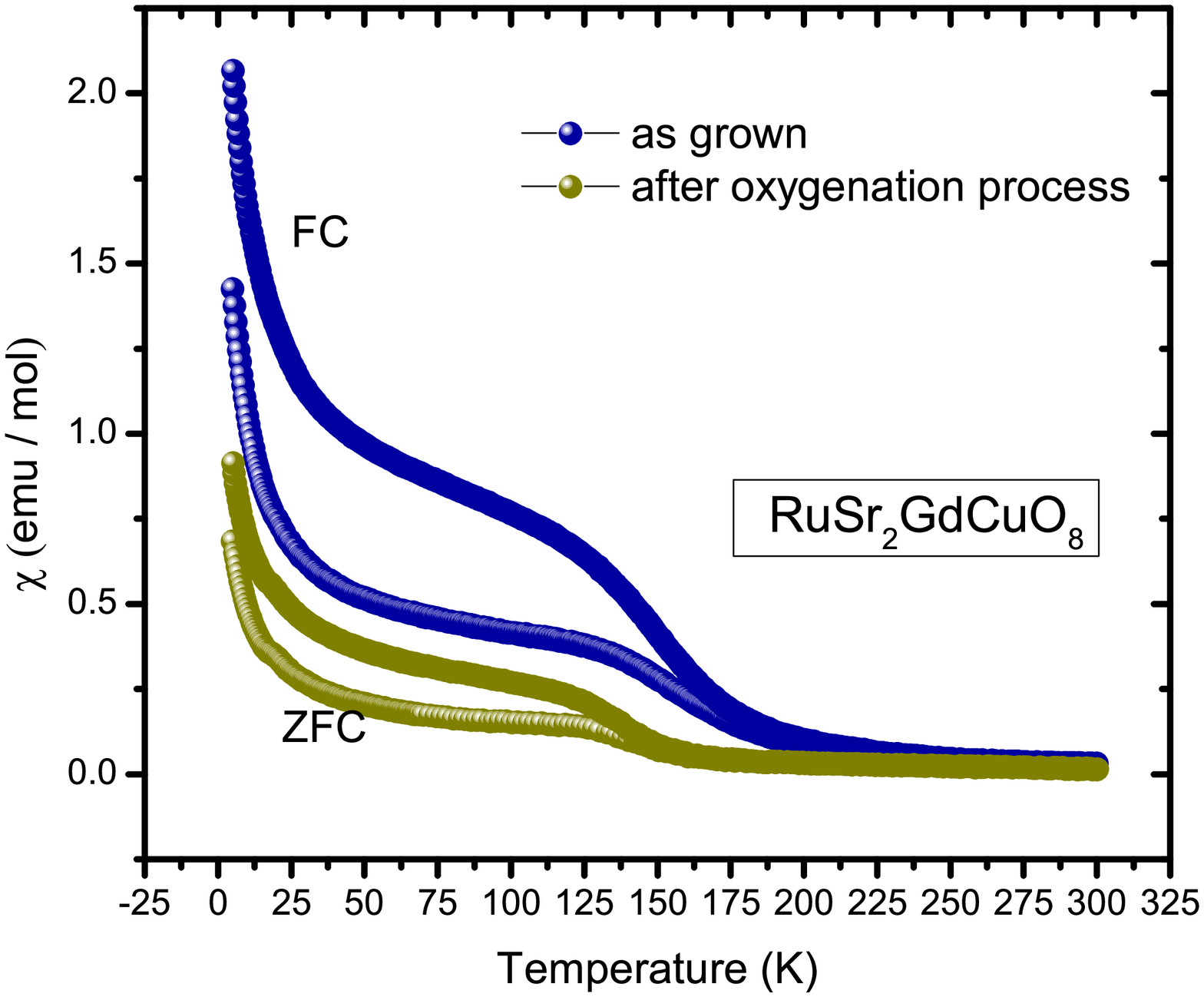}\quad
\includegraphics[width=11cm]{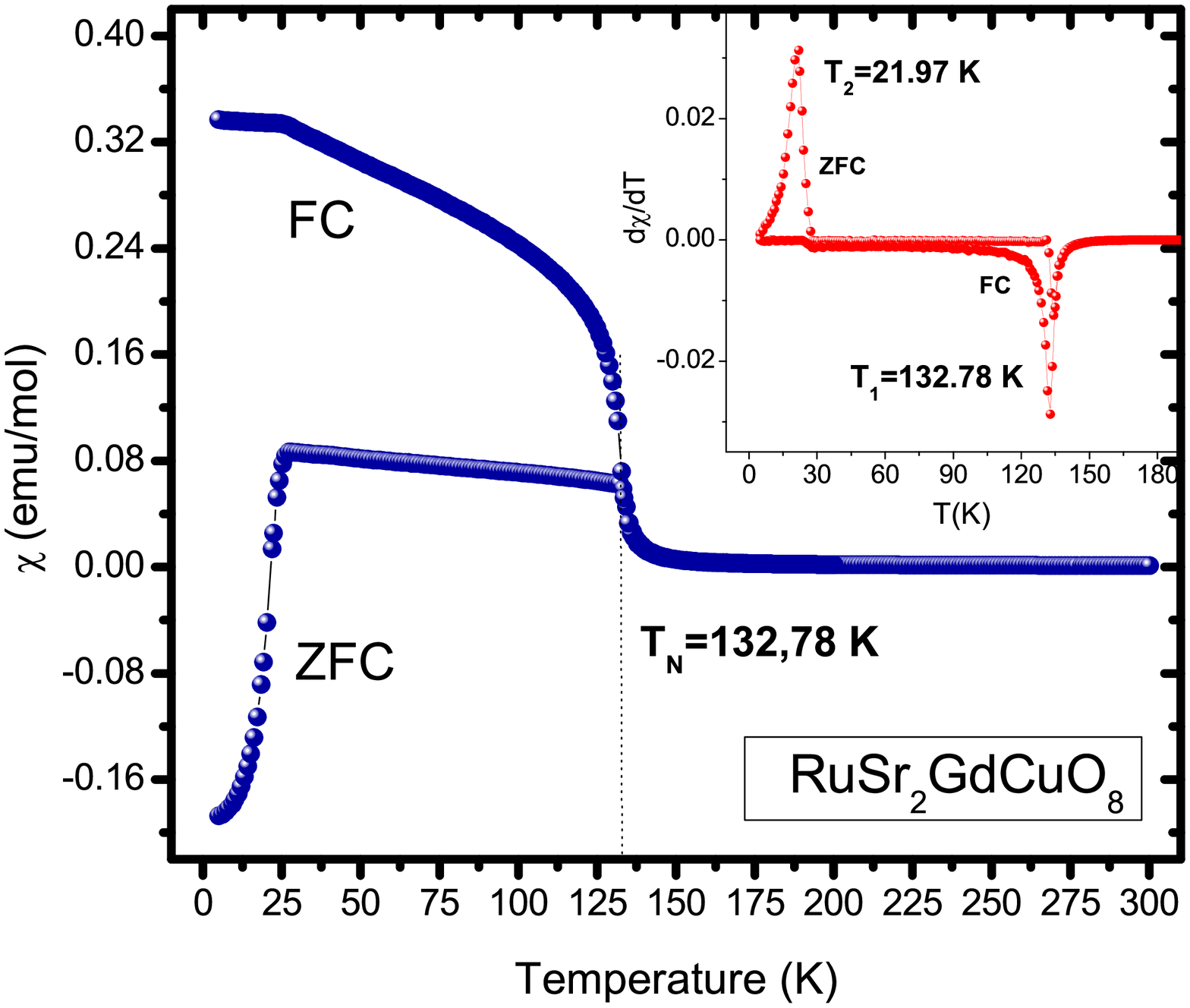}
\caption{}
\label{Fig.2}
\end{figure}

In order to obtain information about the influence of Re on the Ru magnetic lattice, susceptibility measurements were performed. Figure \ref{Fig.2}(a) shows the temperature dependence of the normalized dc susceptibility for the x = 0.00 sample with applied field H = 100 Oe, before and after 20 hours of oxygen treatment. This sample was fabricated to reproduce the pure sample properties at first.  It can be seen that the system exhibits a magnetic transition, attributed to long-range ferromagnetic coupling of the Ru sublattice, at 151 K for the non-oxygenated sample and 139.63 K after oxygen treatment, with a difference of $\Delta_N =$ 11 K.

One of the hypotheses for the origin of this weak ferromagnetic component (WFM) in ruthenocuprates is the rotation of RuO$_6$ octahedra, and the canting of Ru moments via the Dzyaloshinsky-Moriya mechanism. This proposal made by Jorgensen et al. \cite{Jorgensen} states that while dominant ordering of the Ru sublattice is of the antiferromagentic type G, with the easy axes oriented perpendicularly to the layers \cite{Chmaissem,Yelon}, the complete sub-system is slightly tilted due to the Jhan-Teller effect, resulting in a net ferromagnetic component along the $ab$ plane.

On the other hand, the Meissner state is not observed, and under 30 K there is a Curie-Weiss behaviour attributed to the augmented paramagnetism of the sample. After 120 hours of oxygenation, Figure \ref{Fig.2}(b), a superconductor response is induced, with a diamagnetic signal under 22 K, characteristic of the Meissner state. In this case, rising due to Gd ions is not evident, with susceptibility magnitude near constant under 21.9 K.

All the doped samples underwent an oxygenation process at the same time as the pure sample, showing a qualitatively similar behaviour, with irreversibility in the ZFC and FC branches. However, only the 3 \% Re and 6 \% Re samples exhibited a resistivity transition, as explained in section \ref{transport}. The samples did not exhibit evidence of a bulk Meissner state, even after prolonged oxygen treatment. On the other hand, 9 \% Re and 12 \% Re samples did not show noticeable alteration between as-grown and oxygenated samples, so just the oxygenated sample curve is shown. Since a remarkable difference between the susceptibility results before and after oxygen treatment was detected only for the two samples with a resistivity transition, this shows that the emergence of superconductivity strongly affects the magnetic properties of the sample.

When the behaviour of the rhenium-cuprates was compared with the Sr$_2$GdRu$_{1-x}$Re$_x$O$_y$ samples \cite{Corredor}, it was found that for x=0.03-0.06 it is totally the opposite: the magnetic ordering temperature diminishes in rhenium-cuprates instead of increasing (see Table \ref{Table1}). On the other hand, features like reentrance at low temperatures remain the same as in the double perovskites. This reflects the strong paramagnetic contribution of Gd ions, causing the apparent absence of the Meissner state. Also, the T$_N$ of samples with x=0.09-0.12 increases, in accordance with the double perovskites.

\begin{figure}[ht]
\centering
\includegraphics[height=5.5cm,keepaspectratio]{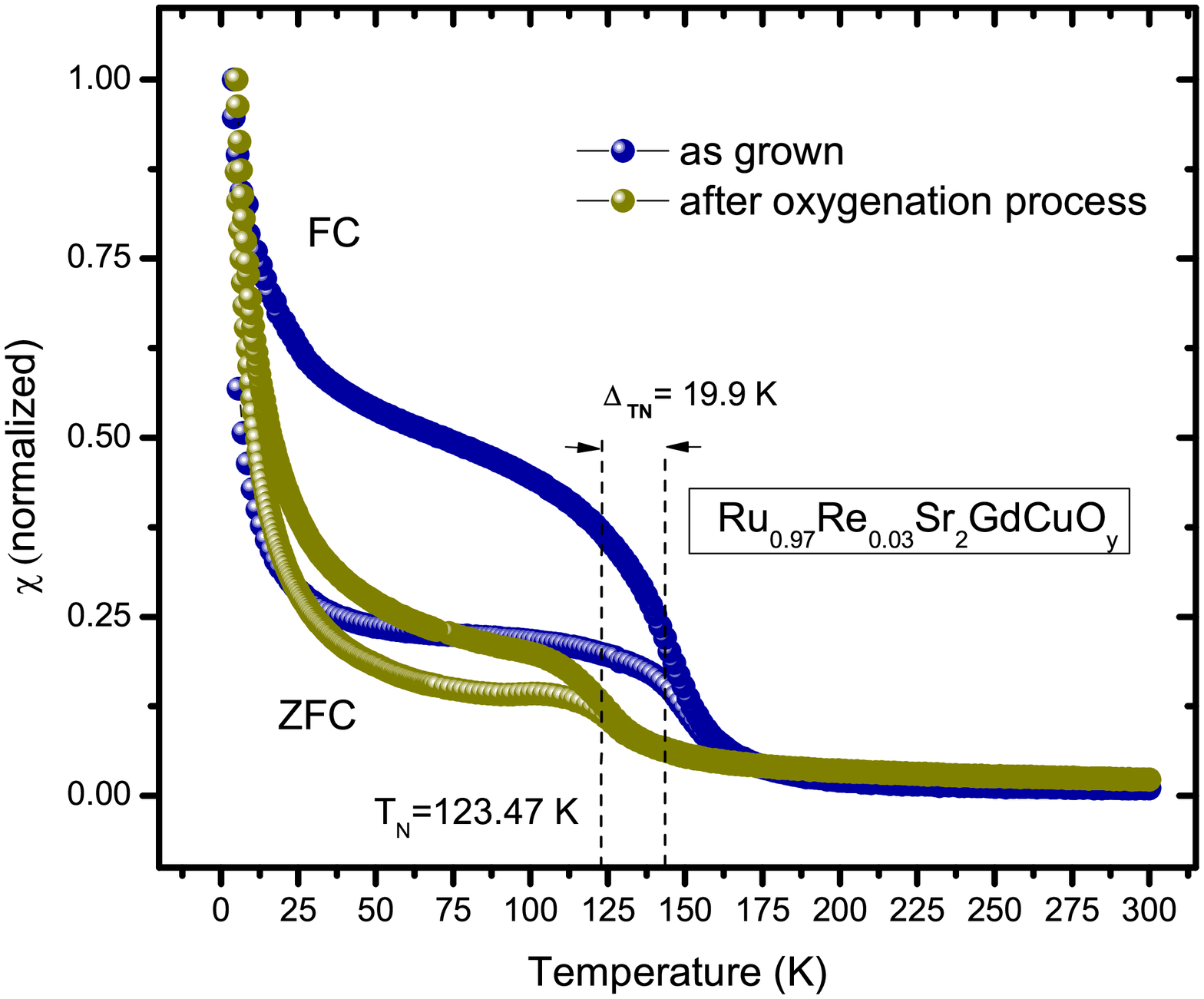}\quad
\includegraphics[height=5.5cm,keepaspectratio]{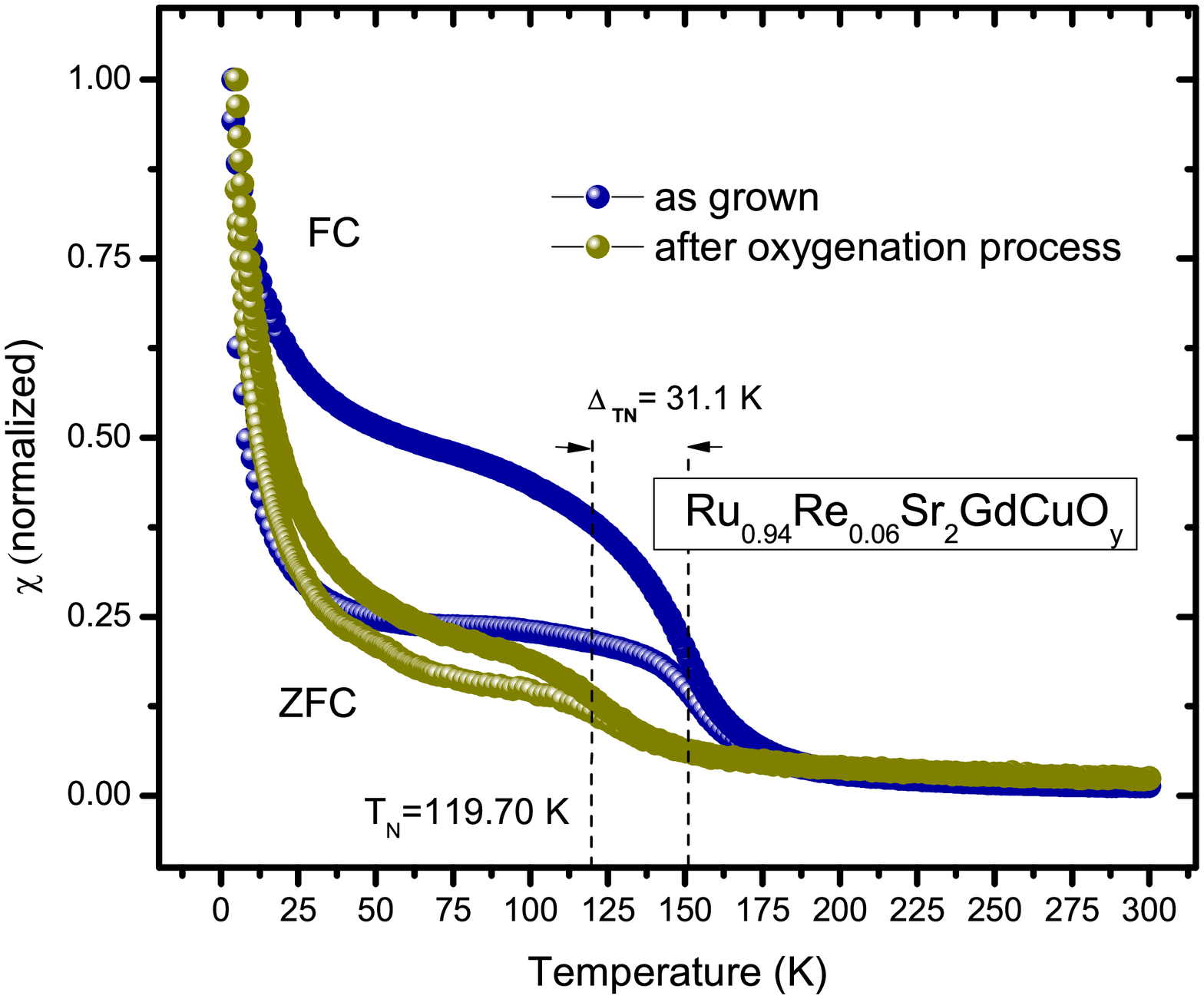}\quad
\includegraphics[height=5.5cm,keepaspectratio]{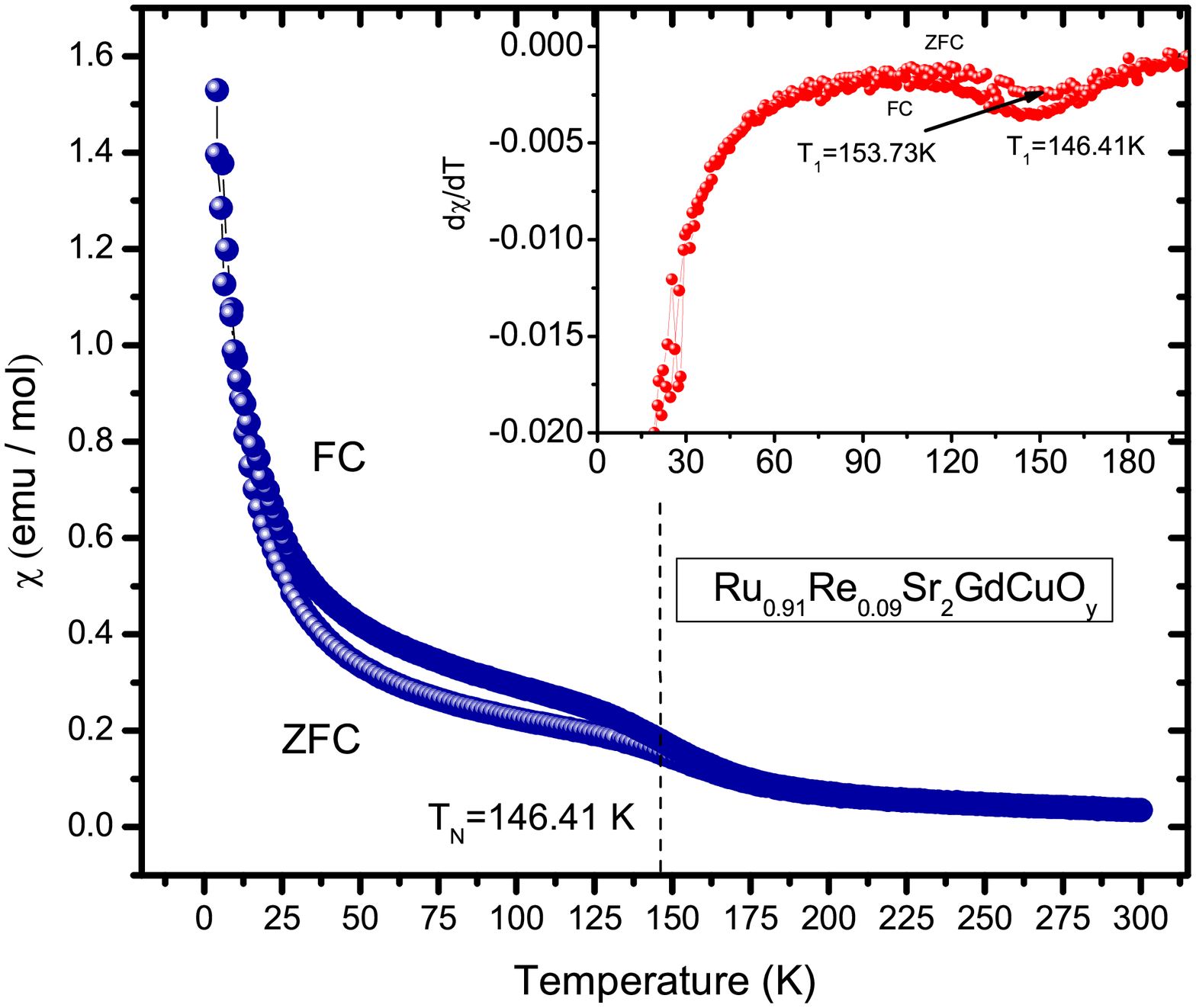}\quad
\includegraphics[height=5.5cm,keepaspectratio]{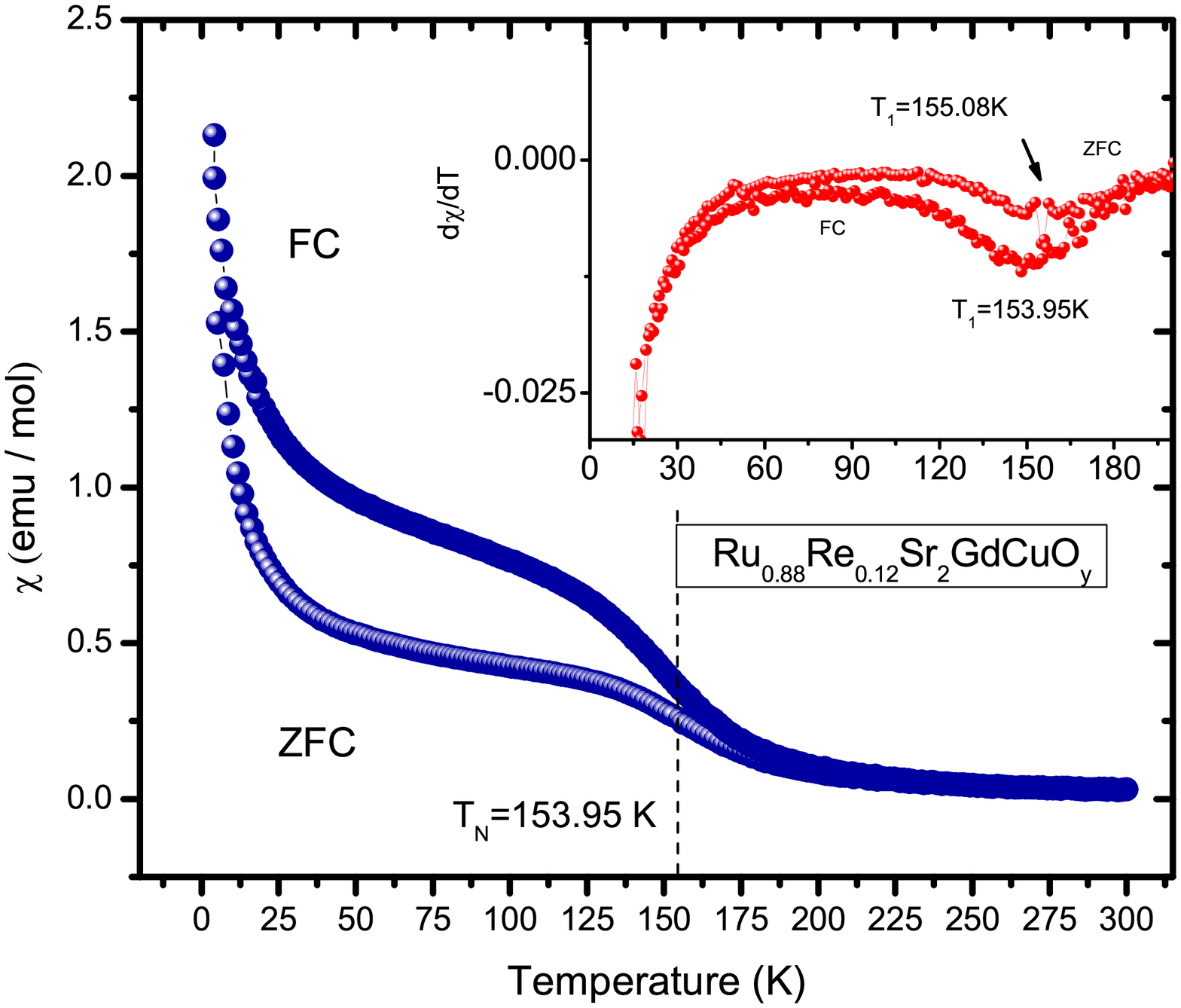}
\caption{}
\label{Fig.3}
\end{figure}

Figure \ref{Fig.3} shows the special behaviour for a 9 \% Re sample, in the same way as was observed for the precursor perovskite Sr$_2$GdRu$_{1-x}$Re$_x$O$_y$. Regarding the magnetic ordering temperature, when starting the doping with Re, the T$_N$ exhibited a small decrease, which then rose with the doping level. However, this scheme has an exception at 9 \% Re: ZFC and FC branches approach each other noticeably, even overlapping at temperatures under 10 K. This effect is similar to the application of strong magnetic fields, which results in the suppression of the irreversibility \cite{Jurelo}. Reduction in the irreversibility suggests a lower cationic disorder, as well as a decrease in the long range weak ferromagnetic order \cite{Klamut2}, since the irreversibility is attributed to the Ru sublattice. Other evidence of this is the reduced value of the coercive field, when compared with the other samples in the series.

Some reports on other ruthenocuprates have linked the increase in the superconductor critical temperature T$_c$ to a decrease in the magnetic ordering temperature, T$_N$. This has been observed in both doping with holes (Cu$^{2+}$) and with heterovalent charge substitutions (Nb$^{5+}$, Sn$^{4+}$) in the Ru sub-lattice, in the same way as in the substitution of (Gd$^{3+}$) for ions like (Ce$^{4+}$) \cite{Klamut3}. This reflects an increase in the transfer of holes to the CuO layers, as well as a reduction of the magnetic order in the RuO$_2$ layers.

In our case, by adding Re to the structure, there exists the possibility of a mixed valence Re$^{5+} (5d - t_2^{2g}$, S= 1) /  Re$^{4+} (5d - t_3^{2g}$, S= 3/2), in this way increasing the number of holes in the CuO layers, besides modifying the magnetic moment of the material. Isovalent substitution (Re$^{5+}$) in principle does not cause doping or hole extraction, but it substantially changes the overlapping of orbital d (Ru/Re) and 2p (O) when 4d of Ru is substituted for 5d of Re. However, this should be considered a mixed valence Ru$^{4+} (4d - t_4^{2g}$, S= 1) /  Ru$^{5+} (4d - t_3^{2g}$, S= 3/2) which would lead to a doping with holes in the CuO$_2$ according to the relation Ru$^{5+}$ + Cu$^{2+} \rightarrow$ Ru$^{5-\delta}$ + Cu$^{2+\delta/2}$ \cite{Casini}.

\begin{figure}[ht]
\centering
\includegraphics[width=12cm,keepaspectratio]{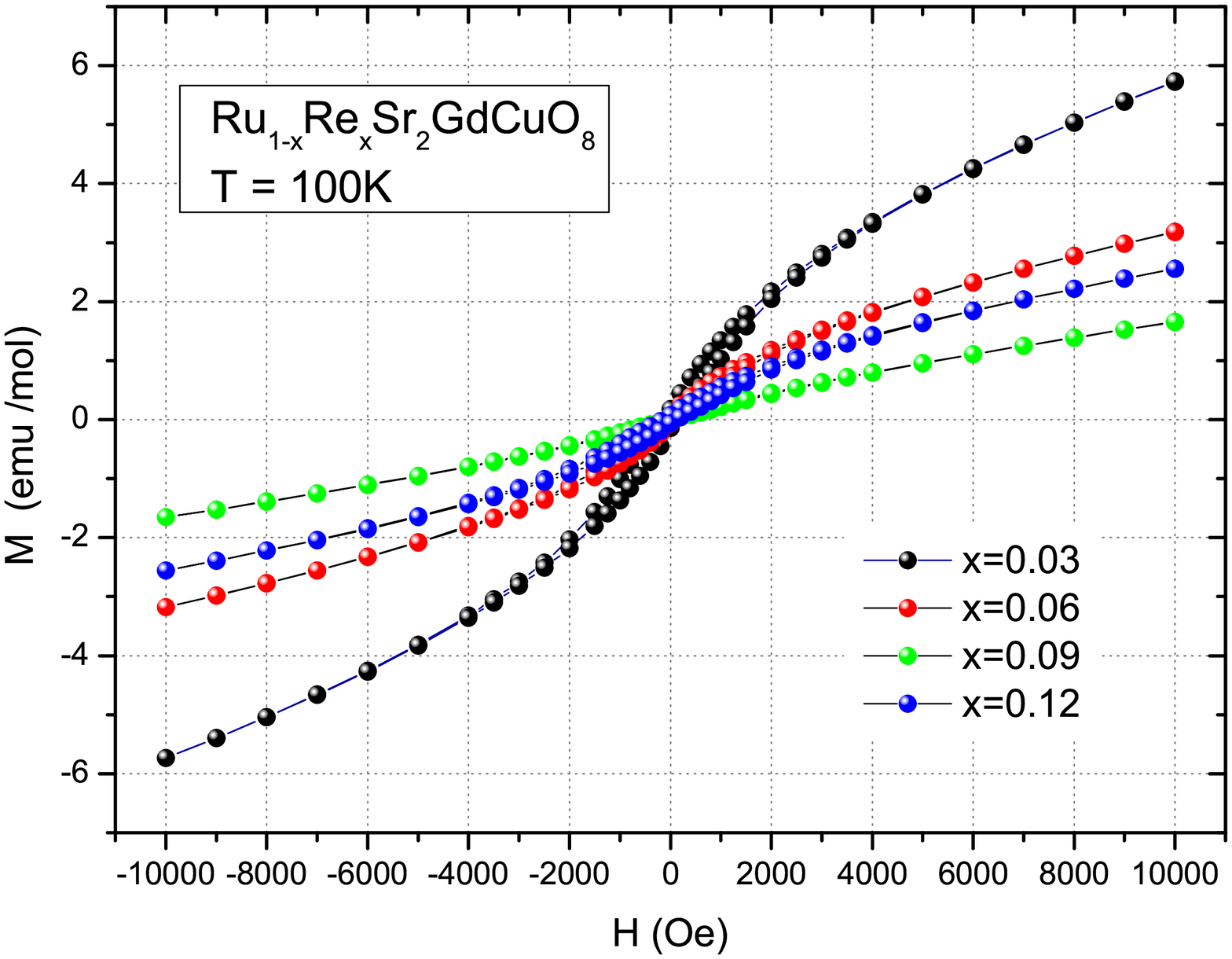}\quad
\includegraphics[width=12cm,keepaspectratio]{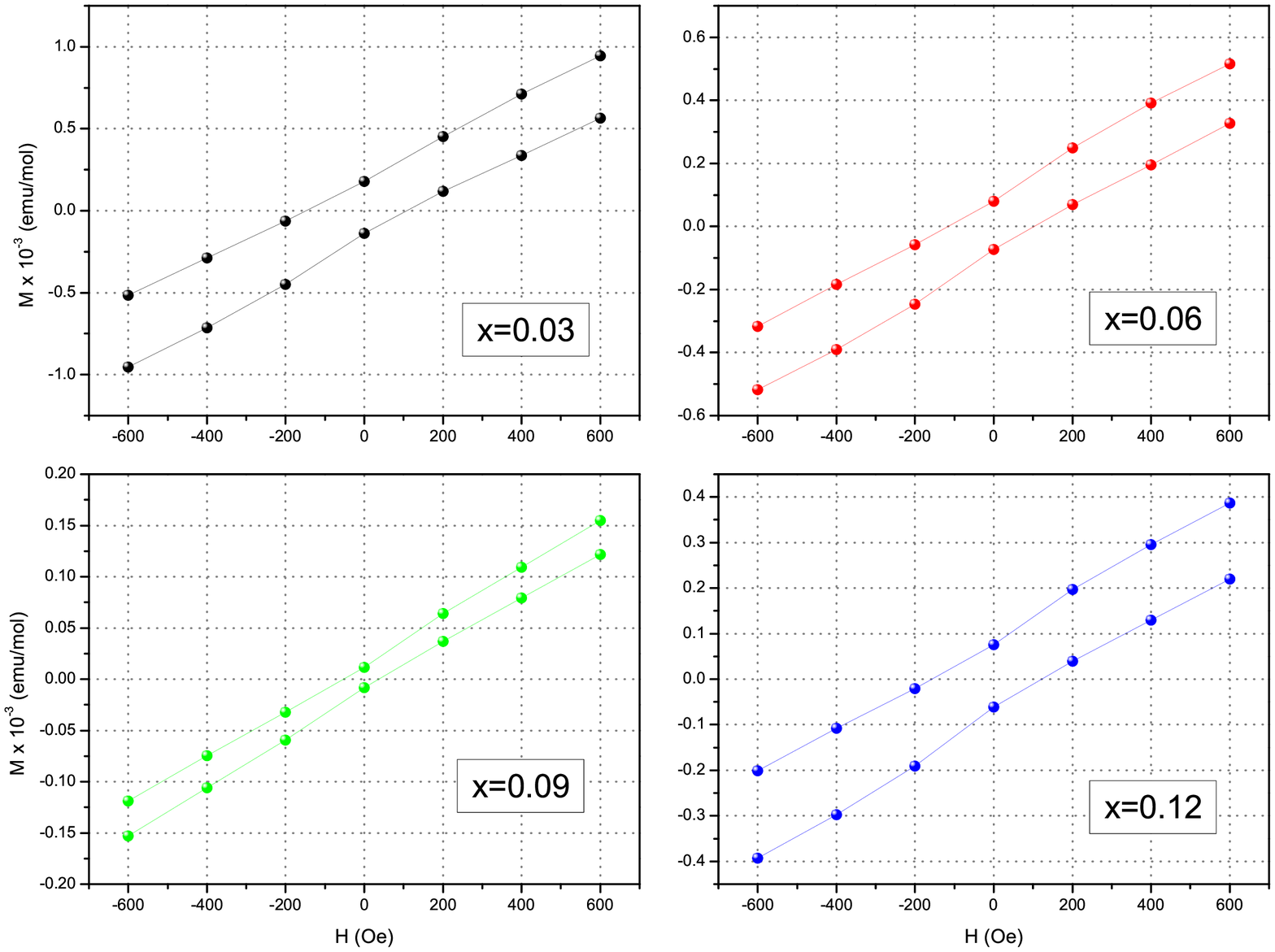}
\caption{}
\label{Fig.4}
\end{figure}

Magnetisation vs applied field measurements at 100 K (under magnetic transition temperature) are shown in Figure \ref{Fig.4}. These measurements revealed a weak ferromagnetic component, with a clear hysteresis loop. Saturation is not reached even for fields until 5 T, which matches a globally antiferromagnetic system, as expected from the Jorgensen model. Graphs are shown in the 1T range for easier observation.

Coercive fields become lower when Re is added to the structure for the 3 \% and 6 \% samples, the same as for the SGReO perovskites. However, the magnetic ordering temperature behaviour is totally the opposite, diminishing instead of increasing. On the other hand, the same parameters for the non-superconducting 9 \% and 12 \% samples behave exactly in the same way as SGReO. This strongly suggests that Re continues favouring antiferromagnetism, in consequence decreasing the weak FM component. Nevertheless, interactions related to the appearance of superconductivity prevent the establishment of AFM ordering at high temperatures, leading to a reduction of the T$_N$ only for the superconducting samples, remaining the same in the other ones. This indicates a close relation between the magnetic lattice (formed by RuO$_2$ layers) and the superconductor lattice (CuO$_2$ lattice), within the two-lattice quasi 2D picture.

Notice that the metamagnetism present in the precursor perovskites SGReO is not present in rhenium-cuprates, indicating on the one hand, the absence of the perovskite phase traces in the samples and on the other that interactions between Ru and Ru/Re ostensibly change from perovskite to ruthenocuprate, leading to very different magnetic properties.

\subsection{Transport}
\label{transport}

\begin{figure}[ht]
\centering
\includegraphics[width=11cm,keepaspectratio]{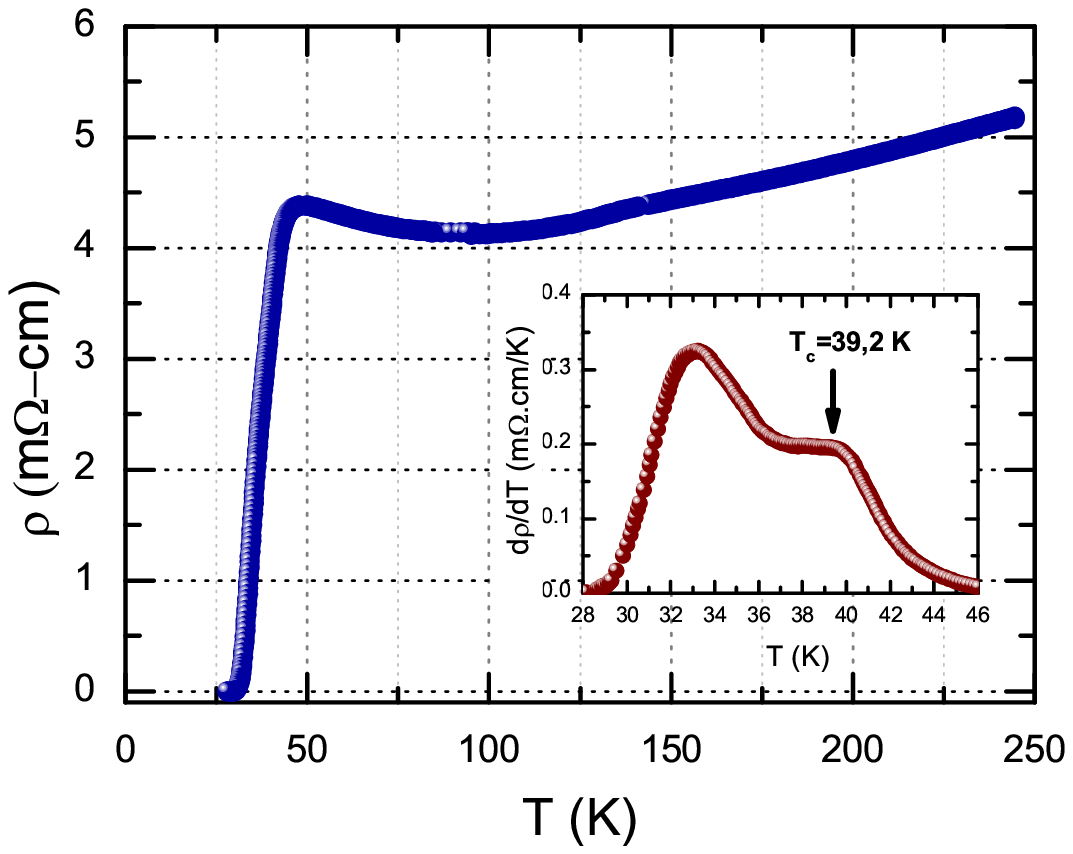}\quad
\includegraphics[width=11cm,keepaspectratio]{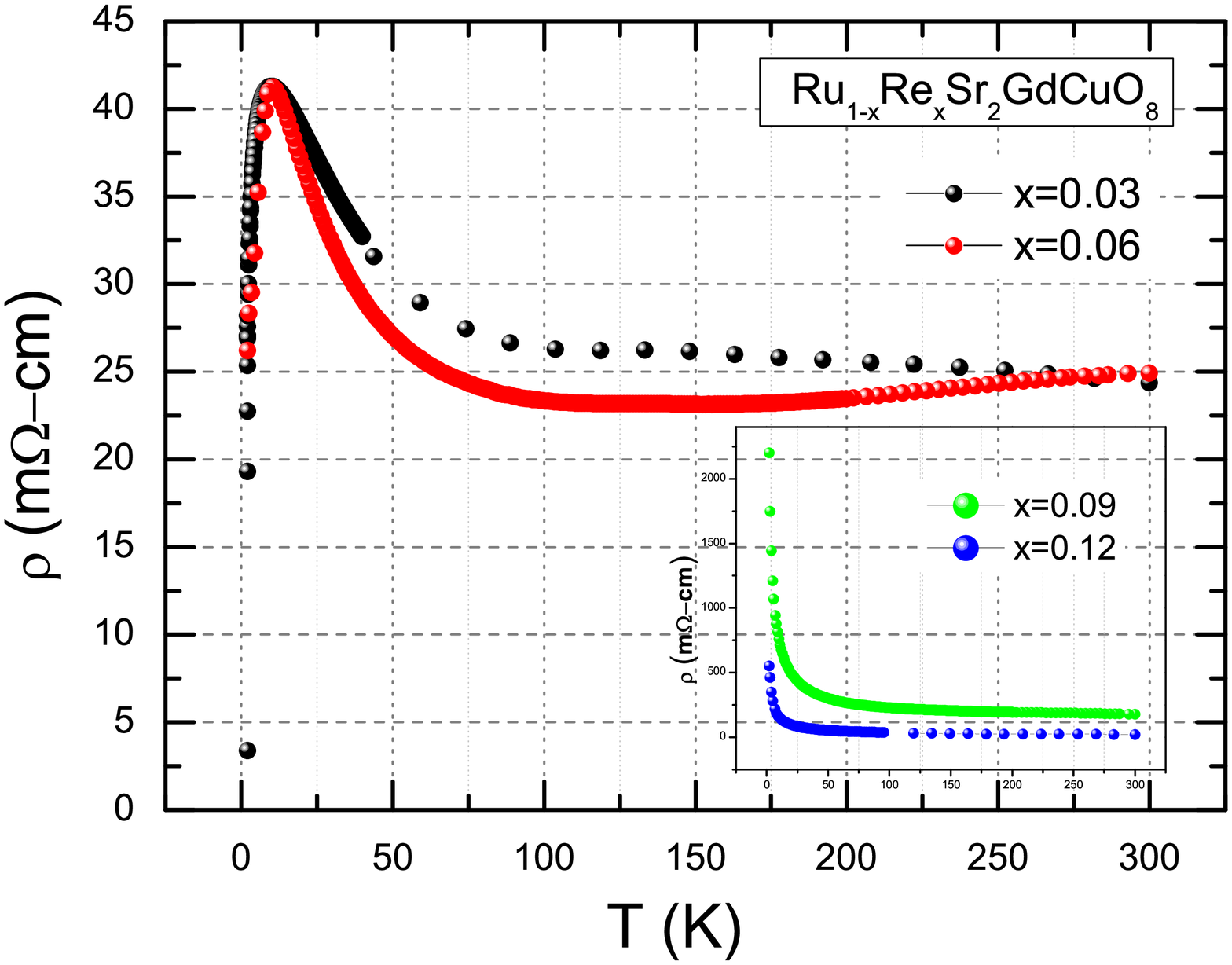}
\caption{}
\label{Fig.5}
\end{figure}

In order to show that magnetic properties participate actively in the conduction mechanisms, resistivity as a function of temperature measurements were performed. The pure oxygenated sample exhibited a linear behaviour in the normal region, followed by a clear superconductor transition at 39.2 K (first derivative criterion), Figure \ref{Fig.5}. When compared with other superconductor cuprates, ruthenocuprates exhibit a larger transition width, in this case approximately 20 K. The derivative peak at 33.1 K indicates the intergranular transition, resulting from the coupling via Josephson and proximity effects.

On the other hand, the doped samples exhibited two kinds of behaviour: for 3 \% and 6 \% doping levels, a semiconductor-type feature can be observed, as well as a drop in resistivity at 9.96 K and 9.86 K, respectively. For 9 and 12 \%, such a transition disappears, retaining the semiconductor-type behaviour at low temperatures. This could be attributed to Re, which would affect the electron coupling with its internal magnetisation, causing the disappearance of the resistivity transition.

The strong contrast between doped ruthenocuprates and 1212-Gd, even for samples with the same oxygen treatment, reflects the strong correlation between the magnetic Ru lattice and the Cu lattice. On the one hand, both the transition temperature and the resistivity behaviour strongly depend on the synthesis process, as is well known. In this type of compounds, strong anisotropy is one of the factors that results in a broad transition, contrary to the abrupt one for other cuprates.

Another scenario is that large transitions could be higher in polycristalline samples, where structural defects such as granularity, grain frontiers, vacancies, etc., influence the spatial modulation of the order parameter on a scale comparable to the coherence length. This enlargement in ruthenocuprates is also related to the magnetic ordering that exists in the Ru sublattice and the possible presence of a spontaneous vortex phase (SVP) \cite{Leviev,Felner2,Tokunaga}. In light of these possibilities, the presence of Re in the sample could cause increased internal magnetisation, which would prevent it from reaching the Meissner state, to such a level of observing just the beginning of the transition, without obtaining the zero resistance state. Also, the 3 \% and 6 \% samples showed a minimum in the graph near the magnetic ordering temperature, as a strong deviation from linearity in the normal state, which could be a consequence of Kondo effect, related to the existence of a localized magnetic moment in the Ru spin lattice. It can be noticed that for the 9 \% and 12 \% samples, with no resistivity transition, this feature is not observed.

\begin{figure}[ht]
\centering
\includegraphics[height=5cm,keepaspectratio]{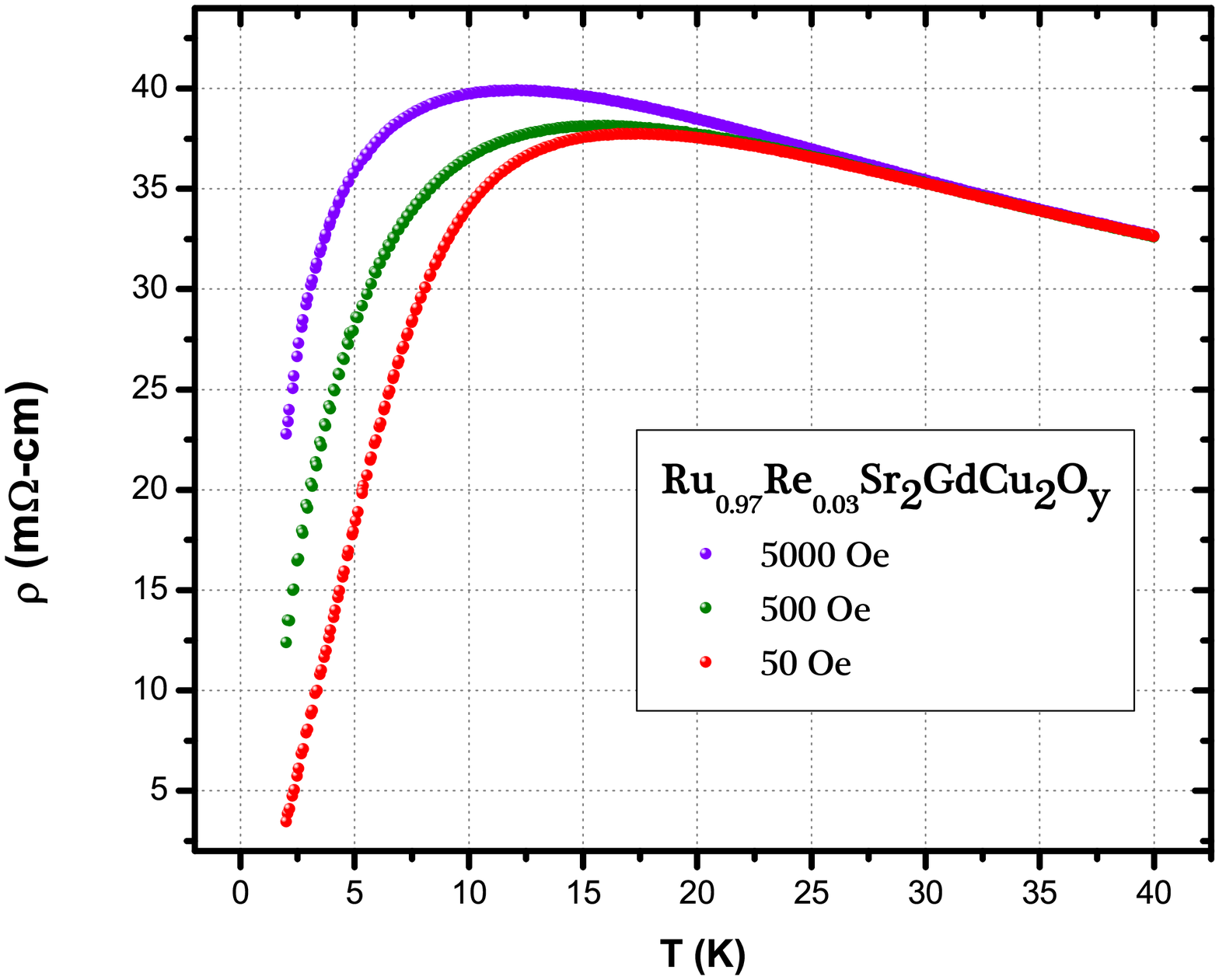}\quad
\includegraphics[height=5cm,keepaspectratio]{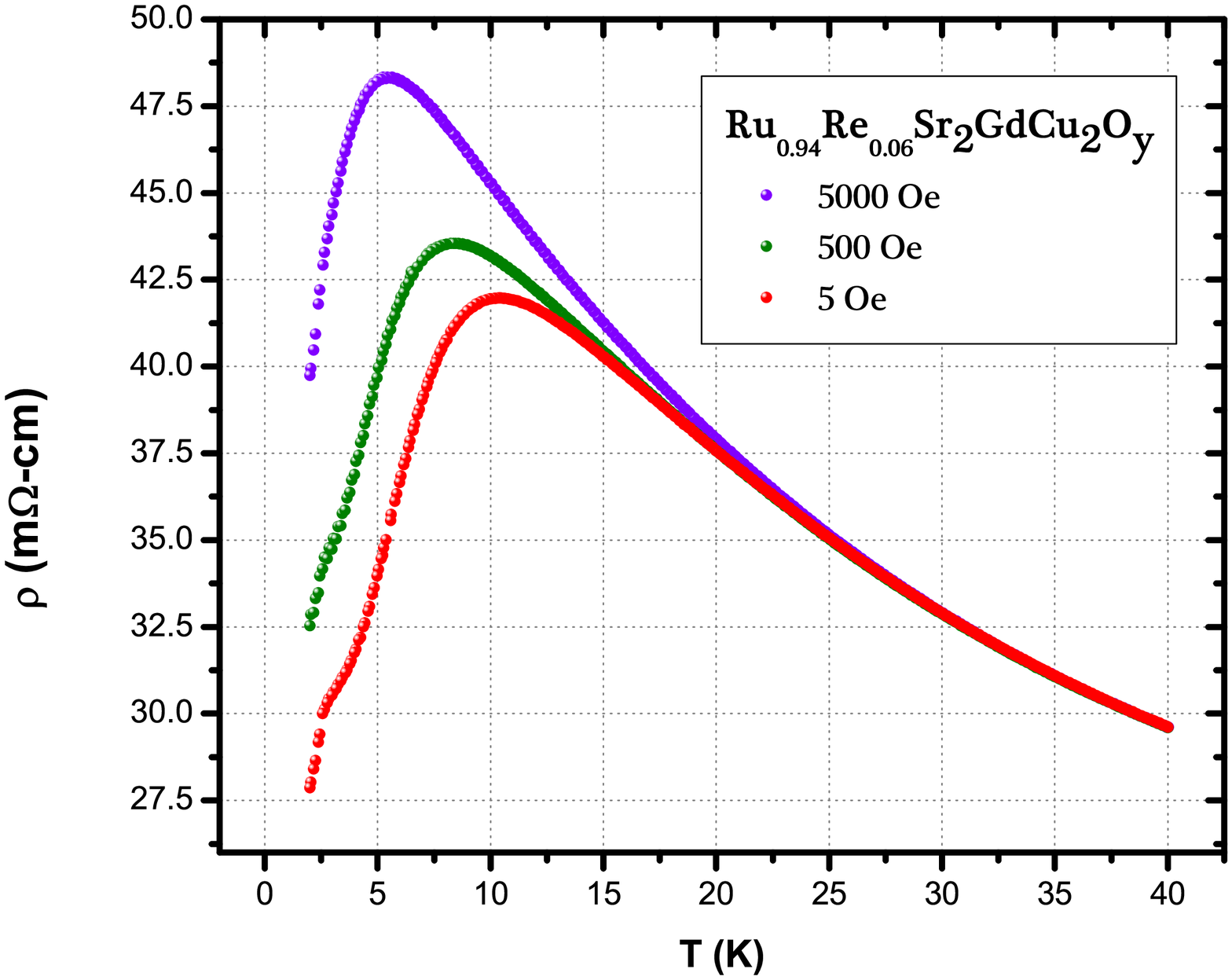}\quad
\includegraphics[height=5cm,keepaspectratio]{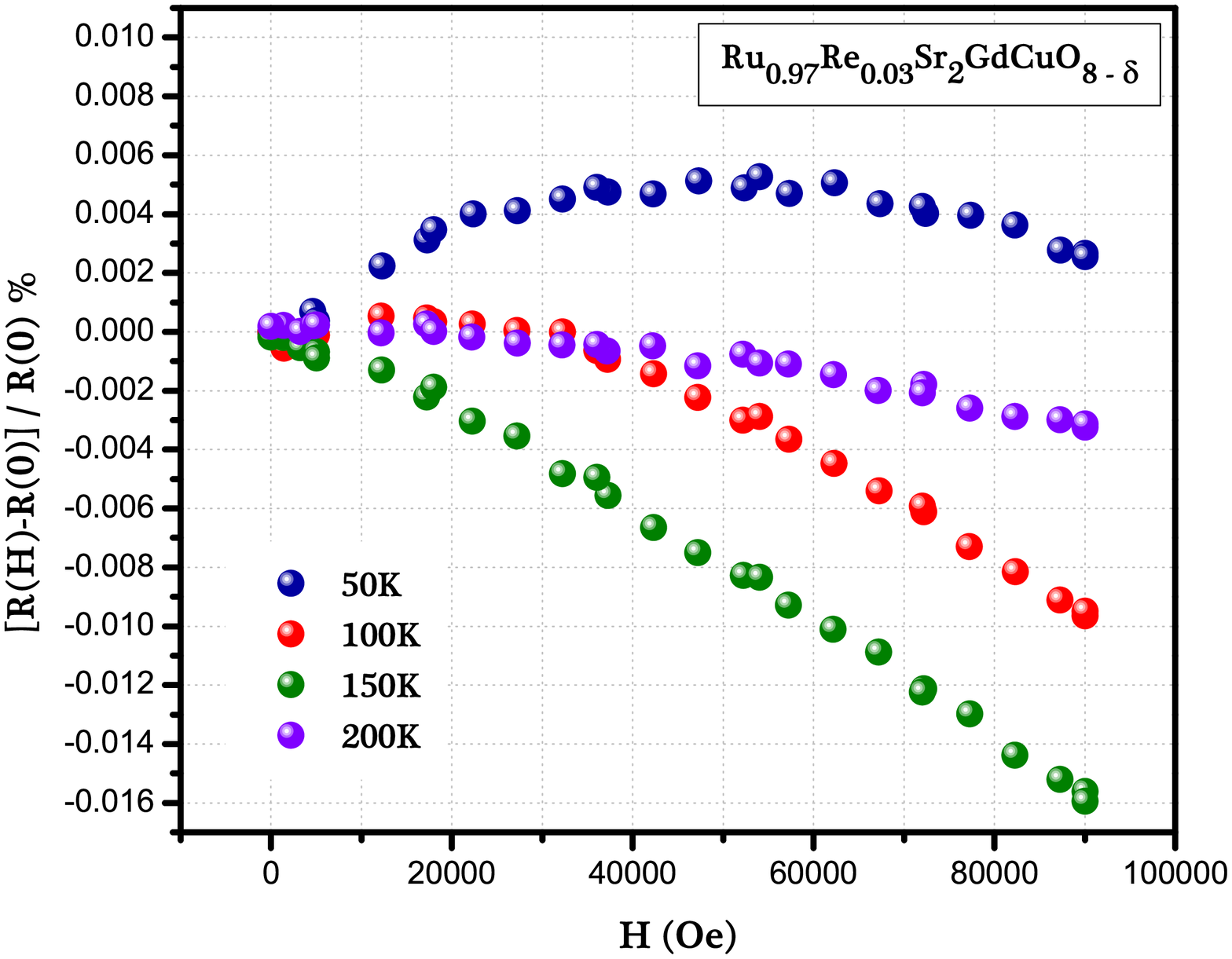}\quad
\includegraphics[height=5cm,keepaspectratio]{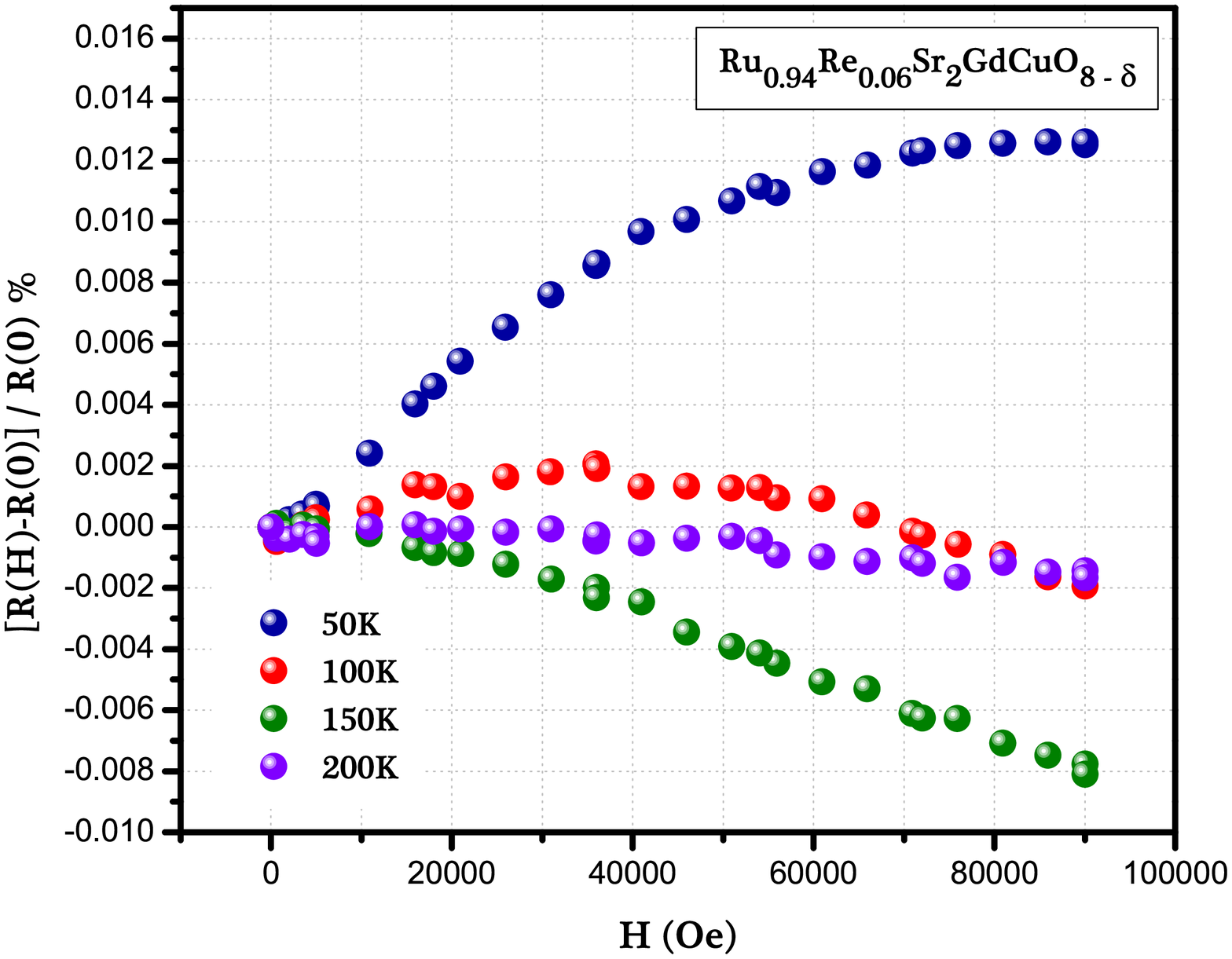}
\caption{}
\label{Fig.6}
\end{figure}

When resistivity measurements as a function of temperature under an applied magnetic field were performed (fields up to 10000 Oe, Figure \ref{Fig.6}), a progressive enlargement of the transition was observed, together with an increase in the residual resistivity while increasing the magnetic field. There was no noticeable influence close to T$_c$, near the intragranular transition.

Even when the curves were not completed within the measurement range (low temperatures), there was a tendency to get closer as the magnetic field increased. These results agree with previous reports with respect to Ru-1212, where this effect was observed for much higher fields (8-12 T) \cite{Chen}. This dependence is quite different from that observed for other high temperature cuprates, governed by a strong vortex motion at low temperatures. Neither is the observed behaviour consistent with a strong flux pinning in a conventional 3D system. These results suggest that the vortex dynamics of these compounds could be intrinsically different from the High Tc cuprates.

Magnetoresistance (MR) measurements were performed, as a tool to provide information on the interaction between charge carriers and magnetic moments. Figure \ref{Fig.6} shows these measurements at several temperatures. It can be observed that for temperatures above magnetic ordering (T$_N\simeq$119-139 K), the MR exhibits just a slight variation, showing a tendency to increase as the temperature increases. A tendency toward linear behaviour can be observed for T=150 K, while at T=100 K, below the T$_N$, the MR shows positive values under 3-7 T, which can be attributed to the positive contribution to the dispersion in the presence of an antiferromagnetic ordering. An applied magnetic field will try to destroy the antiferromagnetic ordering: as a consequence, it will cause an increase in resistivity. On the other hand, a decrease in resistivity could exist due to the suppression of spin fluctuations near the T$_N$ under the application of a magnetic field. Bringing these effects together, we would observe the behaviour of T=100 K curves.

For T=50 K, closer to the beginning of the resistivity transition, the MR rises taking positive values over the whole applied field range, with a clear tendency to decrease as the applied field increases, indicating that the magnetic field significantly influences the charge transport. The antiferromagnetic ordering is completely consolidated, and a stronger field would be necessary to destroy it. In none of the samples the ordering is destroyed before 9 T, nevertheless, the sample with 3 \% Re showed a tendency for the MR to decrease, which was not observed for 6 \% Re. It must be taken into consideration that the MR magnitude for 3 \% Re is lower, so it could be easier to destroy the ordering in this case.

Especially for the 3 \% Re sample, the MR magnitude tends to a maximum at approximately 5.5 T at T=150 K, above the magnetic ordering temperature. This feature could be due to the fact that spin fluctuations, and in consequence the MR, increase as the temperature approaches the T$_N$ from higher temperatures. On the other hand, under the T$_N$ an antiferromagnetic ordering has already been established at zero field, and in this way no change in the charge carriers is expected at the T$_N$, diminishing the dispersion of charge carriers \cite{Klamut4}. In summary, for low applied fields the MR is positive below the magnetic ordering temperature, and negative for higher temperatures: this suggests a behaviour dominated by the interaction between charge carriers and the magnetic moments.

\section{Summary and conclusions}

We investigated the effect of the dilution through Re-doping of the magnetic Ru sub-lattice on the superconducting properties of the well-known Ru-1212, obtaining the new ruthenocuprate compound Ru$_{1-x}$Re$_x$Sr$_2$GdCu$_2$O$_y$ for 3 \%, 6 \%, 9 \% and 12 \% Re. For the synthesis of the material a new procedure was implemented, which at the same time produced the new double perovskite Sr$_2$GdRu$_{1-x}$Re$_x$O$_y$, previously characterized. We took advantage of this to compare the double perovskite properties with the ruthenocuprate ones, in order to evaluate whether their magnetic properties are similar or the ruthenocuprate structure exhibits a unique behaviour.

The magnetic characterisation showed a noticeable difference between results before and after oxygen treatment only for the two
samples with a resistivity transition (3 \% and 6 \% Re), showing that the emergence of superconductivity strongly affects the magnetic properties of the sample. Re presence in the sample could cause increased internal magnetisation, which would prevent it from reaching the Meissner state, as observed in the magnetisation vs temperature results. Further, Re would affect the electron coupling, causing the disappearance of the resistivity transition for higher doping levels.

When comparing the magnetic behaviour of the rhenium-cuprates with the Sr$_2$GdRu$_{1-x}$Re$_x$O$_y$ perovskites, it was found that they are totally different. The tendencies of the magnetic ordering temperature and the coercive field do not correlate in the same way with the doping increase, while features like reentrance at low temperatures are similar. Additionally, the metamagnetism present in double perovskites is totally absent in the ruthenocuprates, confirming on the one hand that that our samples were free of SGReO traces, and on the other hand, that ruthenocuprates have very particular properties.

The magnetoresistance measurements suggested a behaviour dominated by the interaction between the charge carriers and the magnetic moments, while magnetisation vs applied field loops showed a weak ferromagnetic component. These loops did not reach saturation even for fields up to 5 T, which is consistent with a globally antiferromagnetic system, as expected from the Jorgensen model.

We showed the relation that exists between RuO$_2$ and CuO$_2$ lattices, concluding that magnetism plays an important role in the conduction mechanisms, and the dynamics of superconductor behaviour influence the magnetic response of the studied materials. Through further research, neutron diffraction will give more information about the magnetic ordering established, as well as NMR or X-ray absorption characterisations, which are proposed to establish the exact Ru/Re valences.

\section*{Acknowledgments}

This work was partially supported by Colombian agency COLCIENCIAS, Patrimonio Aut\'{o}nomo Fondo Nacional de Financiamiento para la Ciencia, la Tecnolog\'{i}a y la Innovaci\'{o}n Francisco Jos\'{e} de Caldas (Contract RC-No. 0850-2012), Brazilian CNPq, and FACEPE (APQ-0589-1.05/08). L.T.C. would like to thank CAPES/FACEPE for PNPD financial support.

\newpage

\begin{table}[h]\footnotesize
\centering
\begin{tabular}{cccccccc}
\toprule
{\bf Sample} & {\bf{\textit{a,b}} (\AA)} & {\bf{\textit{c}} (\AA)} & {$\mathbf{\chi^2}$} & {\bf R$_F$} & {\bf R$_{wp}$} & {\bf T$_N$} & {\bf (+O$_2$) T$_{N}$}\\\midrule
x=0.00 & 3.845(3) & 11.571(5) & 0.97 & 0.11 & 0.07 & 151.00 & 139.63\\
x=0.03 & 3.834(7) & 11.548(3) & 1.09 & 0.11 & 0.08 & 143.45 & 123.47\\
x=0.06 & 3.835(7) & 11.531(0) & 1.13 & 0.10 & 0.08 & 150.98 & 119.70\\
x=0.09 & 3.837(9) & 11.493(8) & 1.14 & 0.11 & 0.09 & 146.49 & 146.41 \\
x=0.12 & 3.836(7) & 11.515(4) & 1.05 & 0.10 & 0.09 & 153.90 & 153.95\\
\bottomrule
\end{tabular}
\caption{{\small Lattice parameters, refinement reliability factors, and magnetic parameters of Ru$_{1-x}$Re$_x$Sr$_2$GdCu$_2$O$_y$.}}\label{Table1}
\end{table}

\newpage

\centerline{\bf FIGURE CAPTIONS}
\vspace{3cm}

{\bf Figure 1.} (a) XRD pattern for Ru$_{1-x}$Re$_x$Sr$_2$GdCu$_2$O$_y$ family; (b) XRD pattern for x=0.03 sample. Symbols
correspond to the experimental data and the continuous line is the obtained by Rietveld refinement. The bottom curve represents the
difference between the experimental and calculated patterns.
\bigskip

{\bf Figure 2.} Susceptibility curves for RuSr$_2$GdCu$_2$O$_8$ (pure sample) (a)before and after 20 hours oxygen treatment, and (b) after 120 hours oxygen treatment.
\bigskip

{\bf Figure 3.} Susceptibility curves for the Ru$_{1-x}$Re$_x$Sr$_2$GdCu$_2$O$_y$ family before and after oxygen treatment.
\bigskip

{\bf Figure 4.} (a) Magnetisation as a function of applied field for doped samples at T=100 K. (b) Low field region, showing the presence of a coercive field due to a weak ferromagnetic component.
\bigskip

{\bf Figure 5.} Resistivity as a function of temperature for (a) pure RuSr$_2$GdCu$_2$O$_8$ and (b) doped Ru$_{1-x}$Re$_x$Sr$_2$GdCu$_2$O$_y$ samples.
\bigskip

{\bf Figure 6.} (a,b) Resistivity as a function of temperature for the different applied fields, and (c,d) magnetoresistance for x=0.03 and x=0.06 samples.

\newpage

\newpage

\newpage

\newpage

\newpage

\newpage

\end{document}